\documentclass[letterpaper, 10 pt, conference]{ieeeconf}  % Comment this line out if you need a4paper

\IEEEoverridecommandlockouts                              % This command is only needed if 
\usepackage{amsmath,amssymb,amsfonts}
\usepackage{algorithmic}
\usepackage{graphicx} 
\usepackage{textcomp}
\usepackage{xcolor}

\usepackage[hidelinks]{hyperref}
\usepackage{enumerate}
\usepackage{multirow}
\usepackage{gensymb}
\usepackage{cite}

\usepackage{booktabs}
\usepackage{colonequals}

\usepackage[font=footnotesize]{caption}
\usepackage{subcaption}

\usepackage{siunitx} % display and formatting of units
\usepackage[capitalize]{cleveref}
\crefname{figure}{Fig.}{Figs.}
\usepackage{cuted}
\usepackage{flushend}
\usepackage{accents}

\usepackage{tablefootnote}

\newcommand{\set}[2]{\left\{#1\;\middle|\;#2\right\}}
\newcommand{\bmat}[1]{\begin{bmatrix}#1\end{bmatrix}}

\newcommand{\unbar}[1]{\underaccent{\bar}{#1}}
\newcommand{\defeq}{\colonequals}

\def\BibTeX{{\rm B\kern-.05em{\sc i\kern-.025em b}\kern-.08em
    T\kern-.1667em\lower.7ex\hbox{E}\kern-.125emX}}

\newcommand{\cross}[1]{{#1}^{\times}}
\renewcommand{\top}{\mathsf{T}} % prettier symbol for transpose
\newcommand{\tr}[1]{{#1}^{\top}}
\newcommand {\R}{\mathbb{R}}
\newcommand{\norm}[1]{\|#1\|}
\newcommand{\normm}[1]{\left\|#1\right\|}

\DeclareMathOperator{\diag}{diag}

\newtheorem{theorem}{\textbf{Theorem}}

\newtheorem{definition}{\textbf{Definition}}
\newtheorem{remark}{\textbf{Remark}}

\newcommand{\dd}{\mathrm{d}}
\renewcommand{\epsilon}{\varepsilon}

\title{\LARGE \bf
Spacecraft Attitude Control Under Reaction Wheel Constraints Using Control Lyapunov and Control Barrier Functions
}

\author{Milad Alipour Shahraki and Laurent Lessard% <-this % stops a space
	\thanks{M. Alipour Shahraki is with the Department of Electrical and Computer Engineering,
		Northeastern University, Boston, MA 02115, USA
		{\tt\small alipourshahraki.m@northeastern.edu}}%
	\thanks{L. Lessard is with the Department of Mechanical and Industrial Engineering,
		Northeastern University, Boston, MA 02115, USA
		{\tt\small l.lessard@northeastern.edu}}%
}

\begin{document}

\maketitle
\thispagestyle{empty}
\pagestyle{empty}

\begin{abstract}
This paper introduces a novel control strategy for agile spacecraft attitude control, addressing reaction wheel-related input and state constraints. An optimal-decay control Lyapunov function quadratic program stabilizes the system and mitigates chattering at low sampling frequencies, while control barrier functions enforce hard state constraints. Numerical simulations validate the method's practicality and efficiency for real-time agile spacecraft attitude control.
\end{abstract}

\section{Introduction}

Agile spacecraft require rapid and precise attitude control to execute critical tasks such as high-resolution imaging and on-orbit servicing. These systems must perform large-angle maneuvers while operating within strict constraints on reaction wheels (RWs) torque and angular momentum. Notable examples include the Pleiades, OrbView, and WorldView satellite systems \cite{cao2016time}.

While research on unconstrained spacecraft attitude control is extensive, addressing RW torque and angular momentum limitations remains a significant challenge in practical implementations. Methods like sliding mode, backstepping, and adaptive control are known for accuracy and robustness but often fail to handle input or state constraints effectively. We now provide a brief survey of some relevant control approaches.

A classical and efficient approach to handling input constraints is via soft constraints or saturation functions \cite{alipour2022practical,zou2019finite}. However, soft constraints require fine-tuning to avoid constraint violations and saturation functions complicate the dynamics and can lead to longer settling times, which is undesirable for rapid maneuvers.

Linearized models have been used in optimization-based methods such as Model Predictive Control (MPC) for spacecraft attitude control \cite{zagaris2018model}. However, spacecraft dynamics' nonlinearity, especially in large-angle maneuvers, can be problematic. Nonlinear MPC (NMPC) \cite{lee2017geometric} is an alternative, but is computationally intensive, limiting its use for agile spacecraft. MPC-based methods also struggle with robustness against uncertainties and disturbances.

Indirect optimal control computes an optimal policy by solving the Hamilton–Jacobi–Bellman (HJB) equation, but this is done offline and is challenging due to nonlinear dynamics and constraints. Inverse optimal control offers a tractable alternative to solving the HJB equation by identifying a control law with a meaningful cost functional \cite{krstic1998inverse}. Control Lyapunov function (CLFs) further simplify solving this problem \cite{krstic1998inverse, luo2005inverse}. Based on Sontag's formula, this approach has been used for spacecraft attitude control \cite{pukdeboon2012control}, though input constraints are often omitted. CLFs also enable stabilization via point-wise min-norm optimization when formulated as a quadratic program (QP) \cite{freeman1995optimal}.

CLF-QP-based approaches systematically handle input constraints. A variant, the rapidly exponentially stabilizing CLF (RES-CLF) \cite{ames2014rapidly}, has been extended to spacecraft attitude control, focusing on input constraints \cite{li2016extended}. However, RW angular momentum constraints remain unaddressed. Additionally, CLF-QPs with fixed CLF decay rates (e.g., RES-CLF-QP) may cause input chattering at low sampling frequencies \cite[Fig.~6]{reher2020inverse}, \cite{morris2013sufficient}. While increasing the sampling frequency mitigates this problem, it also increases computational burden, making real-world implementation impractical.

Control Barrier Functions (CBFs) translate state constraints into input constraints, enabling the rigorous study of system safety by ensuring set invariance. CLFs and CBFs can be combined in a QP, where the CLF constraint is slacked to ensure feasibility. This leads to a sequence of QPs solved at each timestep to determine control actions \cite{ames2016control}.

Zero-order-hold control and robust sampled-data CBFs were used for spacecraft reorientation under pointing and avoid-zone constraints in \cite{breeden2023autonomous}, considering input but not RW angular momentum constraints. A Control Lyapunov-barrier function (CLBF), combining CLFs and CBFs through nonlinear control, was developed in \cite{romdlony2016stabilization} and applied to spacecraft attitude control in \cite{wu2021attitude}, but input constraints were not incorporated, and the method faced criticism for theoretical inconsistencies \cite{braun2020comment}.

This paper presents a chatter-free, computationally efficient controller for agile spacecraft attitude stabilization under RW-related input and state constraints. The contributions of this paper are as follows:
\begin{enumerate}[1.]
    \item Unlike previous works on spacecraft attitude control \cite{alipour2022practical, li2016extended}, the proposed approach encodes both RW-related input and state constraints as hard constraints via a CLF-CBF-QP controller, offering a computationally efficient alternative to MPC-based methods \cite{lee2017geometric}.
    \item Unlike prior works that rely on inverse dynamics \cite{reher2020inverse} for the CLF-QP chattering problem, the proposed approach introduces a dynamically varying weight for the CLF decay rate to mitigate the chatter at low sampling frequencies. The controller also balances control effort and settling time through parameter tuning.
    \item In prior works, controller designs were typically validated through numerical simulations, which showed absolute performance but did not quantify room for improvement. In our paper, we use a \emph{Pareto plot}, which visualizes performance trade-offs and compares proposed controllers to a baseline family of optimal policies. 
\end{enumerate}
The remainder of the paper is structured as follows:  \cref{sec:Preliminaries} covers relevant preliminaries.  \cref{sec:Problem Statement} outlines the problem statement.  \cref{sec:Controller Design} details the controller design.  \cref{sec:Simulation and Results} provides comparative simulations, and \cref{sec:Conclusion and Future Work} concludes the paper with future research directions.

\section{Preliminaries}\label{sec:Preliminaries}

Consider a nonlinear \emph{input-affine} control system
\begin{equation}\label{eq:affine nonlinear system}
	\dot{x} = f(x) + g(x)u, \quad y = h(x),	
\end{equation}
where $x \in \R^n$ is the state, $u \in U\subseteq\R^m$ is the control input, and $y \in \R^\ell$ is the output. In addition, $f: \R^n \to \R^n$ and $g: \R^n \to \R^{n \times m}$ are Lipschitz continuous. We have the following definition for system \eqref{eq:affine nonlinear system}:
\begin{definition}[Relative degree \cite{xiao2021high}] The relative degree of a (sufficiently) differentiable function $h$ with respect to system \eqref{eq:affine nonlinear system} is the number of times we need to differentiate it along the dynamics
of \eqref{eq:affine nonlinear system} until  $u$ explicitly appears.\looseness=-1
\end{definition}

\subsection{Stability via Control Lyapunov Functions (CLF)}\label{subsec:Control Lyapunov Function}

\begin{definition}[CLF \cite{freeman1996control}]\label{def:CLF}
A continuously differentiable function $V: \R^{n} \to \R$ is a
\emph{Control Lyapunov Function (CLF)} for system \eqref{eq:affine nonlinear system}, if it is positive definite and satisfies
\begin{equation}\label{eq:CLF Condition}
    \inf_{u \in U}\;\bigl( L_{f}V(x) + L_{g}V(x)u \bigr) \leq -W(x)
    \quad\text{for all }x,
\end{equation}
where $L_{f}V$ and $L_{g}V$ denote first-order Lie derivatives along $f$ and $g$, respectively and $W$ is a continuous, positive definite function such that $L_{f}V(x) \leq -W(x)$ whenever $L_{g}V(x)=0$. If $V$ is a CLF, define for each $x$ the set\\
$K_{\textup{CLF}}(x) \defeq \bigl\{u \in U \mid L_{f}V(x) + L_{g}V(x)u \leq -W(x)\bigr\}$.
\end{definition}

The function $W$ represents the desired decay rate of the closed-loop $\dot{V}$, with different choices for $W$ leading to different inverse optimal control laws \cite{freeman1996control}. 
\begin{theorem}[See \cite{freeman1996control}] \label{thm:CLF} For system \eqref{eq:affine nonlinear system}, if there exists a CLF $V: \R^{n} \to \R$, i.e., a positive definite function satisfying \eqref{eq:CLF Condition}, then any Lipschitz continuous controller $u$ satisfying $u(x) \in K_{\textup{CLF}}(x)$ for all $x \in \R^{n}$ asymptotically stabilizes the system to the origin.
\end{theorem} 

\subsection{Safety via Control Barrier Functions (CBF)}\label{subsec:Control Barrier Function}

\begin{definition}[CBF \cite{xiao2021high}]\label{def:CBF}
Let $\mathcal{C}$ be the superlevel set of a continuously differentiable function $B: \R^{n} \to \R$, so that
$\mathcal{C} = \{x \in \R^{n} \; | \; B(x) \geq 0\}$.
$B$ is a \emph{Control Barrier Function (CBF)} for system \eqref{eq:affine nonlinear system} if there exists $\alpha \geq 0$ such that
\begin{equation}\label{eq:CBF Condition}
    \sup_{u \in U}\; \bigl( L_{f}B(x) + L_{g}B(x)u \bigr) \geq -\alpha B(x) \quad\text{for all }x \in \mathcal{C}.
\end{equation}
If $B$ is a CBF, define for each $x\in \mathcal{C}$ the set\\
$K_{\textup{CBF}}(x) \defeq \bigl\{u \in U \; \mid \; L_{f}B(x) + L_{g}B(x)u \geq -\alpha B(x)\bigr\}$.
\end{definition}

\begin{theorem}[See \cite{xiao2021high, ames2016control}] \label{thm:CBF}
Let $\mathcal{C}$ be defined as in \cref{def:CBF}. If $B$ is a CBF for system~\eqref{eq:affine nonlinear system}, then any Lipschitz continuous controller $u(x) \in K_{\textup{CBF}}(x)$
has the property that if $x(0)\in \mathcal{C}$, then $x(t)\in\mathcal{C}$ for all $t\geq 0$.
\end{theorem}
The property of $\mathcal{C}$ in \cref{thm:CBF} is also called \emph{forward invariance} \cite{ames2019control}.

\section{Problem Statement}\label{sec:Problem Statement}

\subsection{Spacecraft Attitude Model}

For simplicity, we assume three identical, axially symmetric RWs aligned with the spacecraft's principal axes. This ensures the wheel rotations do not affect the system's moment of inertia. The angular momentum of the RWs $h_{w}$ is calculated as $h_{w}=J_{w} \omega_{w}$,
where $J_{w} = \diag(J_{w1}, J_{w2}, J_{w3})$, and $\omega_{w} \in \R^{3}$ are the RWs moment of inertia and angular velocity, respectively. In this formulation, the spacecraft's angular velocity contribution to the RWs' angular momentum is neglected because it is small compared to the RWs' angular velocity. Although the RWs' angular velocity can be large, it is always bounded, ensuring that $h_{w}$ also remains bounded \cite{cao2016time}.\looseness=-1

Attitude dynamics are given by the equations \cite{zou2020fixed}
\begin{subequations}\label{eq:kindyn}
\begin{align}
	\dot{\sigma} &= M(\sigma) \omega,\label{eq:kinematics}\\
	J\dot{\omega} &= -\cross{\omega} (J \omega + h_w) + u + d,\label{eq:dynamics}\\
	\dot{h}_w &= -u,\label{eq:rwd}
\end{align}
\end{subequations}
where $\sigma \in \R^3$ is the attitude representation using Modified Rodrigues Parameters (MRPs), $J \in \R^{3\times3}$ and $\omega \in \R^{3}$ are the inertia matrix and angular velocity of the spacecraft, expressed in the body-fixed frame, $d \in \R^{3}$ is the external disturbance torque and $u \in \R^{3}$ is the control torque. 
The notation $\cross{x} \in \R^{3 \times 3}$ denotes the cross-product matrix operator.
Additionally, $M(\sigma) \defeq \frac{1}{4}\bigl((1 - \tr{\sigma}\sigma)I_{3} + 2\cross{\sigma} + 2\sigma\tr{\sigma}\bigr)$.
\cref{eq:kinematics} captures the kinematics and \eqref{eq:dynamics} and \eqref{eq:rwd} describe the spacecraft and RWs dynamics, respectively.

With the state vector defined as $x = \tr{\begin{bmatrix}\tr{\sigma}&\tr{\omega}&\tr{h}_w\end{bmatrix}}$ and external disturbances neglected, the dynamics \eqref{eq:kindyn} can be written in input-affine form \eqref{eq:affine nonlinear system}, with
\begin{equation}\label{eq:fxgx}
f(x) \defeq \bmat{M(\sigma) \omega \\
		-J^{-1} \cross{\omega}(J\omega + h_w)\\
		0_{3 \times 1}}\!,\;
g(x) \defeq \bmat{0_{3 \times 3} \\
		J^{-1} \\
		-I_{3}}\!.
\end{equation}

\begin{remark}
In this paper, we consider a \emph{rest-to-rest maneuver} where
$x(0) = \tr{\bmat{\tr{\sigma_0} & \tr{0} & \tr{0}}}$ and $x_{\textup{final}} = 0$. If the final state is nonzero, then instead of \eqref{eq:fxgx}, we should use the so-called \emph{relative error equations} \cite{zou2020fixed}.    
\end{remark}

\section{Controller Design}\label{sec:Controller Design}

In this section, we design a stabilizing controller based on the CLF formulation \eqref{eq:CLF Condition}. We then discuss how formulating the CLF with a fixed decay rate in a QP can cause input chattering at low sampling frequencies and propose a fix. Finally, we encode the RWs angular momentum constraint (state constraint) using the CBF concept \eqref{def:CBF} and formulate a CLF-CBF-QP for the spacecraft attitude control system \eqref{eq:kindyn}.
\subsection{CLF-QP Design}\label{subsec:CLF-QP Design}

We begin by summarizing the input-output linearization procedure and applying it to the spacecraft attitude dynamics to form a quadratic CLF. While the focus of this paper is on spacecraft attitude control, the formulations in this subsection can be used to design a CLF for a general nonlinear input-affine dynamical system.
If the system \eqref{eq:affine nonlinear system} has relative degree $r$, and $\ell = m$, then we have
\begin{equation}\label{eq:output_yr}
	y^{(r)} = L_{f}^{r}h(x) + L_{g}L_{f}^{r-1}h(x)u,
\end{equation}
where $L_{f}^{r}h(x)$ and $L_{g}L_{f}^{r-1}h(x)$ are the $r^{\text{th}}$ order Lie derivatives, and $y^{(r)}$ is the $r^{\text{th}}$ time derivative of $y$.
If $L_{g}L_{f}^{r-1}h(x) \in \R^{m \times m}$ is nonsingular for all $x \in \R^n$, then we can apply a control input that renders the input-output dynamics of the system linear:\looseness=-1
\begin{equation}\label{eq:u_mu}
	u(x, \mu) = u^{*}(x) + (L_{g}L_{f}^{r-1}h(x))^{-1} \mu,
\end{equation}
where $u^{*}(x) \defeq -( L_{g}L_{f}^{r-1}h(x))^{-1} L_{f}^{r}h(x)$ is the feedforward term and $\mu \in \R^{m}$ is the \emph{auxiliary input}. Using this control law yields the input-output linearized system $y^{(r)} = \mu$, and we can define a state transformation $x \mapsto (\eta, z)$, with the transverse coordinates $\eta \defeq \tr{\begin{bmatrix}\tr{h(x)} & \tr{L_{f}h(x)} & \cdots & \tr{L_{f}^{r-1}h(x)}\end{bmatrix}}$ and the zero-dynamics manifold $z \in \{x \in \R^n \;|\; \eta = 0\}$. The closed-loop dynamics of the system can then be represented as a linear time-invariant system on $\eta$ and $z$:
\begin{equation}\label{eq:linearized system}
	\dot{\eta} = F \eta + G \mu, \quad \dot{z} = p(\eta, z),
\end{equation}
where $F \!=\! \addtolength{\arraycolsep}{-1pt}\bmat{%
		0 & 1 & \cdots & 0 \\
		\vdots & \vdots & \ddots & \vdots \\
		0 & 0 & \cdots & 1 \\
		0 & 0 & \cdots & 0} \otimes I_{m}$,
 $G \!=\! \bmat{%
		0 \\
		\vdots \\
		0 \\
		1} \otimes I_{m}$, and  $F \in \R^{m r \times m r}$, $G \in \R^{m r \times m}$, and $\otimes$ is the Kronecker product.

It can be verified that $V(\eta) = \tr{\eta}P\eta$ is a CLF for the input-output linearized system \eqref{eq:linearized system}. Defining $\bar{f} \defeq F \eta$, $\bar{g} \defeq G$, we have: $\dot{V}(\eta, \mu) = L_{\bar{f}}V(\eta) + L_{\bar{g}}V(\eta)\mu$, where $L_{\bar{f}}V(\eta) = \tr{\eta} (\tr{F}P + PF) \eta$ and $L_{\bar{g}}V(\eta) = 2\tr{\eta}PG$. Now consider the following quadratic cost:
\begin{equation}\label{eq:cost function J}
    J = \int_{0}^{\infty}\bigl({\tr{\eta}Q\eta} + \tr{u}\bar{R}u\bigr) \,\dd t,
\end{equation}
where $Q \in \R^{m r \times m r}$, $\bar{R}=\nu I_{m} \in \R^{m \times m}$ and $\nu$ is a design variable to penalize the input usage for the initial nonlinear system. Let $\bar{L} \defeq L_{g}L_{f}^{r-1}h(x)$ and $\bar{\mu} \defeq L_{f}^{r}h(x)$. Since the input of the linearized system is $\mu$, we can rewrite \eqref{eq:cost function J} as
\begin{equation}\label{eq:cost function J 2}
    J = \int_{0}^{\infty}\Bigl( \tr{\eta}Q\eta + \tr{(\mu-\bar{\mu})}R(\mu-\bar{\mu}) \Bigr) \,\dd t,
\end{equation}
where $R \defeq \bar{L}^{-\top}\bar{R}\bar{L}^{-1} = \nu(\bar{L}^{-\top}\bar{L}^{-1}) \in \R^{m \times m}$. Under standard stabilizability and detectability assumptions, there is a unique symmetric positive definite solution $P$ to the continuous-time algebraic Riccati equation
\begin{equation}\label{eq:cRiccati}
    \tr{F}P + PF +Q - PGR^{-1}\tr{G}P  = 0.
\end{equation}
Now consider \eqref{eq:CLF Condition}. If we choose $W(\eta)$ as
\begin{align}\label{eq:W(eta)}
    \nonumber &W(\eta) = \sqrt{(L_{\bar{f}}V(\eta))^2 + (\tr{\eta}Q\eta)(L_{\bar{g}}V(\eta))R^{-1}\tr{(L_{\bar{g}}V(\eta))}}\\
    \nonumber &= \sqrt{(\tr{\eta}(\tr{F}P + PF)\eta)^{2}+4(\tr{\eta}Q\eta)(\tr{\eta}PGR^{-1}\tr{G}P\eta)}\\
    &= \tr{\eta}(Q+PGR^{-1}\tr{G}P)\eta,
\end{align}
then the QP based on this formula recovers the standard Linear Quadratic (LQ) optimal control law $\mu = -R^{-1}\tr{G}P\eta$ when the system dynamics are linear, and produces an efficiently computable control strategy when the system is nonlinear and feedback linearizable \cite{freeman1995optimal}.

For our spacecraft attitude control problem, we choose the output $y= h(x) = \sigma$. With this choice, the system has a relative degree of $r = 2$, giving $\eta = \tr{\begin{bmatrix}\tr{\sigma}&\tr{\dot{\sigma}}\end{bmatrix}}$.
\begin{remark}
    The zero dynamics for the spacecraft attitude control problem exist and are stable. For details, see \cite{akhtar2020feedback}.
\end{remark}

\subsection{Addressing Input Chattering}\label{sec:chatter}

The CLF-QP formulation based on \eqref{eq:CLF Condition} with \eqref{eq:W(eta)} forms a point-wise min-norm optimization problem ensuring system stability per \cref{thm:CLF}. However, solving this QP at low sampling frequencies can cause input chattering \cite{morris2013sufficient}, due to the fixed CLF decay rate generating aggressive control signals to maintain stability over longer intervals between updates. 

In \cite{zeng2021safety}, a variable decay weight was introduced for the CBF to ensure the feasibility of the CBF and input constraints in a CBF-QP. Inspired by this, we introduce the variable decay weight $\rho\in[0,1]$ for the decay rate of the CLF in \eqref{eq:CLF Condition} as:
\begin{equation}\label{eq:OD-CLF Condition}
    \inf_{u \in U}\;\bigl( L_{f}V(x) + L_{g}V(x)u \bigr) \leq -\rho W(x).
\end{equation}
We include $\rho$ in the QP as an extra decision variable, which dynamically adjusts the decay rate of the CLF to balance stabilization speed with input smoothness. By varying the decay rate, we prevent aggressive corrections that cause chattering, ensuring smoother inputs without compromising stability.\looseness=-1
\begin{remark}
    In \eqref{eq:W(eta)}, if $R = I$, an alternative choice is
    \begin{equation}\label{eq:RES-CLF W(eta)}
        W(\eta) = \frac{1}{\epsilon}\frac{\lambda_{\min}(Q)}{\lambda_{\max}(P)}V(\eta),
    \end{equation}
    where $\lambda_{\min}$ and $\lambda_{\max}$ denote minimum and maximum eigenvalues and $\epsilon$ is a design variable. This formulation, termed RES-CLF, was first introduced in \cite{ames2014rapidly}. While the variable $\epsilon$ adjusts the decay rate of the RES-CLF, manually tuning it to achieve the optimal settling time without inducing chattering is challenging. Adding a variable decay weight as described above can also help control chattering with a RES-CLF. 
\end{remark}

\subsection{CLF-CBF-QP Design}\label{subsec:CLF-CBF-QP Design}

We now apply the CBF concept to encode box constraints on the RW angular momentum. The constraints are given by $\unbar{h}_{w} \leq h_{w} \leq \bar{h}_{w}$, where
$\unbar{h}_{w} \defeq h_{w, \min}\tr{\begin{bmatrix}1&1&1\end{bmatrix}}$ and
$\bar{h}_{w} \defeq h_{w, \max}\tr{\begin{bmatrix}1&1&1\end{bmatrix}}$.
Define the vector-valued CBFs
$\unbar{B}(x) \defeq h_{w} - \unbar{h}_{w}$
and $\bar{B}(x) \defeq \bar{h}_{w} - h_{w}$.
These CBFs encode the minimum and maximum angular momentum constraints for each RW, respectively.

The Jacobian matrices for these CBFs are given by $\nabla \unbar{B}(x) = -\nabla\bar{B}(x) = \tr{\begin{bmatrix}
			0_{3 \times 3} & 0_{3 \times 3} & I_{3}
	\end{bmatrix}}$.
Therefore, the Lie derivatives with respect to the spacecraft dynamics \eqref{eq:fxgx} are $L_{f}\unbar{B}(x) = L_{f}\bar{B}(x) = 0_{3 \times 1}$ and
$L_{g}\unbar{B}(x) = -L_{g}\bar{B}(x) = -I_{3}$.
The CBF constraints $L_{f}\unbar{B}(x) + L_{g}\unbar{B}(x) u \geq -\alpha \unbar{B}(x)$ and $L_{f}\bar{B}(x) + L_{g}\bar{B}(x) u \geq -\alpha \bar{B}(x)$ can therefore be expressed succinctly as $-\alpha\bar{B}(x) \leq u \leq \alpha\unbar{B}(x)$.

Now using the CLF constraint \eqref{eq:OD-CLF Condition} and the CBF constraint above, we obtain our proposed Optimal-Decay CLF-CBF-QP (OD-CLF-CBF-QP): 
\begin{align}\label{eq:Optimal Decay CLF-QP3}
\underset{u,\rho,\delta}{\min} \quad & \normm{\bar{L}(u-u^{*})}_H^2 + p_{\rho}(1-\rho)^{2} + p_{\delta} \delta^{2} \\
\text{s.t.}\quad & L_{\bar{f}}V(\eta) + L_{\bar{g}}V(\eta) \bar{L}(u-u^{*}) \leq -\rho W(\eta) + \delta, \notag\\
&-\alpha\bar{B}(x) \leq u \leq \alpha\unbar{B}(x),
\; \unbar{u} \leq u \leq \bar{u},
\; \rho \geq 0,\notag
\end{align}
where $H \succ 0$ is a weighting matrix and $\norm{x}_H^2 \defeq \tr{x} H x$.
Our formulation \eqref{eq:Optimal Decay CLF-QP3} includes a weight $p_{\delta} > 0$ for the slack variable $\delta \in \R$, which ensures the CLF constraint is always feasible.
The weight $p_{\rho}>0$ penalizes the variable decay weight $\rho\geq 0$ and we use an associated cost $(1-\rho)^2$ to favor $\rho=1$ (the nominal decay weight).
The input torque bounds are $\unbar{u} = u_{\min}\tr{\begin{bmatrix}1&1&1\end{bmatrix}}$ and $\bar{u} = u_{\max}\tr{\begin{bmatrix}1&1&1\end{bmatrix}}$. 

Our OD-CLF-CBF-QP \eqref{eq:Optimal Decay CLF-QP3} is always feasible due to the slack variable $\delta$ (e.g., let $u=0$), and guarantees safety since state and input constraints are never violated due to \cref{thm:CBF}.
Consequently, $\dot V \leq 0$ is not guaranteed to hold at every timestep. However, we can monitor $\delta$ in real-time and whenever $\delta=0$, we are assured by \cref{thm:CLF} that $x\to 0$.
The main ways we tune our controller are:
\begin{itemize}
    \item Varying the CBF decay rate $\alpha>0$. Increasing $\alpha$ means the controller will be more aggressive and allow itself to get closer to state constraint boundaries.
    \item Varying the input penalty $\nu$ in \eqref{eq:cost function J}. Increasing $\nu$ results in using less input and favors slower maneuvers.
\end{itemize}

\section{Simulation and Results}\label{sec:Simulation and Results}

In this section, we provide an empirical evaluation of our controller and compare its performance to that of alternative control strategies, including an optimal control strategy.
The simulation parameters we used are given in \cref{table:Simulation parameters}.

\begin{table}[ht!]
\renewcommand{\arraystretch}{1.3}
\centering
\caption{Simulation Parameters}\label{table:Simulation parameters}
\begin{tabular}{ll}
\toprule
    Inertia Matrix & $\mathbf{J}=\bigg[\begin{smallmatrix}
        1.8140 & -0.1185 & 0.0275 \\
        -0.1185 & 1.7350 & 0.0169 \\
        0.0275 & 0.0169 & 3.4320
    \end{smallmatrix}\bigg]\,\unit{kg.m^2}$ \\
    RWs Torque Limits & $u_{\max}=-u_{\min}=\qty{0.123}{N.m}$ \\
    RWs Momentum Limits & $h_{w,\max}=-h_{w,\min}=\qty{0.50}{N.m.s}$ \\
    Initial Euler Angles\tablefootnote{Converting Euler angles to MRPs yields:\\$\sigma_0 = \addtolength{\arraycolsep}{-1mm}\tr{\begin{bmatrix} 0.332&-0.614&0.587\end{bmatrix}}$.} (3-2-1) & $\psi_{0}=\tr{\begin{bmatrix}\ang{140}&\ang{20}&\ang{100}\end{bmatrix}}$ \\
    Initial Angular Velocities & $\omega_{0}=\tr{\begin{bmatrix}0&0&0\end{bmatrix}}\,\unit{rad/s}$ \\
    Sampling Frequency & \qty{10}{Hz} (QP solves per second)\\
\bottomrule
\end{tabular}
\end{table}

\subsection{Optimal Control Simulations}

This subsection formulates a family of optimal control problems as a baseline for comparison with the proposed OD-CLF-CBF-QP \eqref{eq:Optimal Decay CLF-QP3}. For different final times $T_{\text{final}} \in (0, 100]$, we solved the nonlinear energy-optimal control problem (OCP) in \eqref{eq:OCP}, establishing a \emph{Pareto Optimal Curve} that trades off control effort and total maneuver time. For OCP simulations, we used the Python Control Systems Library \cite{fuller2021python}, which implements a direct discretize-then-optimize approach using the SLSQP solver to obtain an open-loop optimal trajectory.
For our proposed OD-CLF-CBF-QP and all other QP-based controllers, we used CVXPY with the PROXQP solver \cite{bambade2022prox}.
All simulations were conducted on an Apple M2 Pro chip with 32GB of RAM.

The energy-optimal control problem (OCP) is given by
\begin{align}\label{eq:OCP}
    \underset{u}{\min} \quad & \int_{0}^{T_{\text{final}}}\norm{u}_H^2 \, \dd t \\
	\text{s.t.}\quad & \dot{x} = f(x) + g(x) u, \notag\\
        & x(0) = x_{0}, \; x(T_{\text{final}}) \in \mathcal{T}, \notag\\
        & \unbar{h}_{w} \leq h_{w} \leq \bar{h}_{w}, \;\; \unbar{u} \leq u \leq \bar{u}, \notag
\end{align}
where $x = \tr{\begin{bmatrix}\tr{\sigma}&\tr{\omega}&\tr{h_{w}}\end{bmatrix}}$ and $H = I_{3}$.
We used the terminal set
$
\mathcal{T} = \set{x}{\norm{\sigma}_\infty \leq 0.02 \text{ and } \norm{\omega}_\infty \leq 0.005}
$.
For QP-based controllers, since we cannot explicitly specify the final time, we ran the controller for a long time and then computed  $T_\text{final}$ post hoc using
\[
T_\text{final} = \inf \set{ t \geq 0 }{ x(\tau) \in \mathcal{T} \text{ for all } \tau \geq t }.
\]

\begin{figure}[htp]
	\centering
	\includegraphics{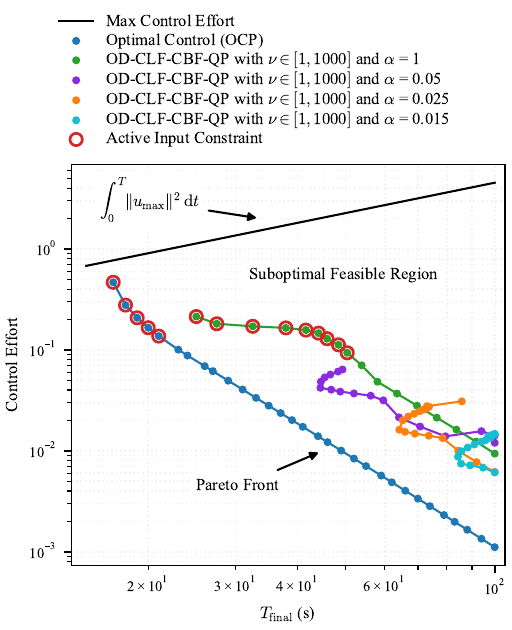}\vspace{-1mm}
	\caption{The Pareto curve, derived from solving the OCP \eqref{eq:OCP} for various final times, is compared to
 the proposed OD-CLF-CBF-QP \eqref{eq:Optimal Decay CLF-QP3} across different tunings of $\nu$ and $\alpha$, with log-scaled $x$-$y$ axes.\vspace{-2mm}}
 \label{fig:pareto_front}
\end{figure}

Results are shown in \cref{fig:pareto_front}. The OCP curve serves as a lower bound indicating the most efficient possible maneuver for a given $T_\text{final}$. Meanwhile, the max control effort line (using $u(t) = u_\text{max}$ for all $t\geq 0$), serves as an upper bound.

In \cref{fig:pareto_front}, we show a variety of tunings of $\nu$ and $\alpha$ for our proposed OD-CLF-CBF-QP to illustrate the trade-off between input energy and total maneuver time.
Increasing $\nu$ (input penalty) favors slower maneuvers by using less input. Likewise, increasing $\alpha$ (CBF decay rate) allows trajectories to get closer to state constraint boundaries, resulting in faster maneuvers. Each tuning is guaranteed to produce feasible QPs and safe trajectories. As we tune the controller, we obtain performances that trace out a trade-off curve roughly parallel to the OCP curve, indicating that control effort is roughly a constant factor worse than optimal.

\subsection{Comparative Simulations}\label{sec:comparative simulations}

We now compare five controllers, again with the simulation parameters from \cref{table:Simulation parameters}. Each controller was tuned manually to achieve good performance and comparable $T_\text{final}$.
\begin{enumerate}[1.]
    \item PD controller with saturation, using $k_p = 0.4$, $k_d = 0.8$. This controller does not require solving a QP.\label{c1}
    \item RES-CLF-QP \cite{ames2014rapidly, li2016extended}, using $Q = I_{6}$, $K_{1} = 0.01\cdot I_3$, $K_{2} = 0.05\cdot I_3$, $\epsilon = 0.2$, $H = I_{3}$, and $p_{\delta} = 100$.\label{c2}
    \item OD-CLF-QP, which is the same as our proposed OD-CLF-CBF-QP controller \eqref{eq:Optimal Decay CLF-QP3}, but without the CBF constraint $-\alpha\bar{B}(x) \leq u \leq \alpha\unbar{B}(x)$. We used $p_{\rho} = 0.1$, $\nu = 10$, and the same $H$, $p_\delta$, and $Q$ as in \cref{c2}.\label{c3}
    \item Our proposed OD-CLF-CBF-QP controller \eqref{eq:Optimal Decay CLF-QP3}, tuned using the parameters of \cref{c3} and $\alpha=0.05$.\label{c4}
    \item OCP \eqref{eq:OCP}. We chose $T_\text{final}$ to match the other controllers. We plotted open-loop performance to serve as a ``best possible'' comparative benchmark.\label{c5}
\end{enumerate}

\begin{figure}[htp]
	\centering
	\includegraphics{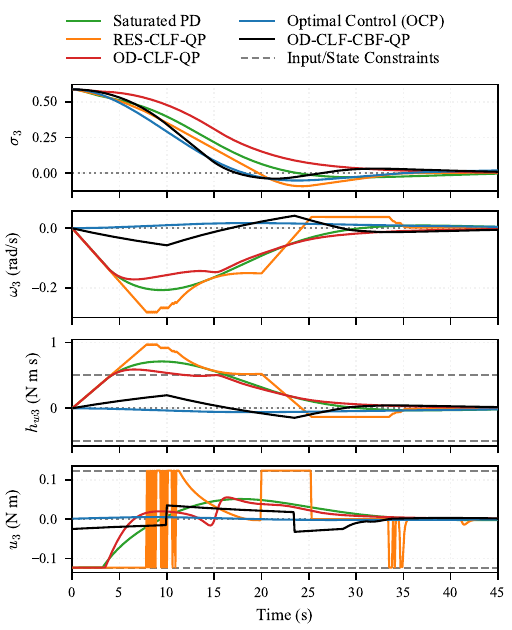}\vspace{-1mm}
	\caption{Representative attitude MRPs, angular velocity, RW angular momentum, and control torque simulation responses for various controllers.
 Only the 3\textsuperscript{rd} components of $\sigma$, $\omega$, $h_w$, and $u$ are shown.}\label{fig:Sim}
\end{figure}

\begin{figure}[htp]
	\centering
	\includegraphics{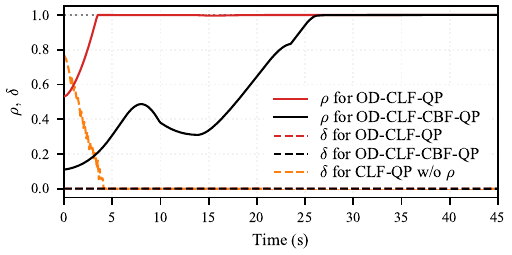}\vspace{-1mm}
	\caption{Optimal decay weight $\rho$ and slack variable $\delta$ simulation response for OD-CLF-QP, and OD-CLF-CBF-QP. Since $\delta \to 0$, we are assured that the trajectory is asymptotically stable.\vspace{-5mm}}\label{fig:rho}
\end{figure}

We note that controllers \ref{c1}--\ref{c3} satisfy the input constraints, but cannot guarantee that state constraints (e.g., the RW constraint on $h_w$) will be satisfied.
Resulting state and input trajectories are shown in \cref{fig:Sim}, the evolution of decay weights and slack variables in \cref{fig:rho}, and a comparison of the total wall clock time (all iterations) is given in \cref{table:Performance comparison}.

\begin{table}[ht!]
\renewcommand{\arraystretch}{1.3}
\centering
\caption{Total wall clock time comparison for different controllers. The corresponding state and input trajectories are shown in \cref{fig:Sim}. }\label{table:Performance comparison}
\begin{tabular}{clcc}
\toprule
    &Control Method & Wall Clock Time & Cost $\int_{0}^{T}\norm{u}^2\,\dd t$ \\
    \midrule
    1.&Saturated PD & $ \qty{0.12}{sec}$ & $0.1892$ \\
    2.&RES-CLF-QP & $ \qty{1.44}{sec}$ & $0.8002$ \\
    3.&OD-CLF-QP & $ \qty{1.92}{sec}$ & $0.1744$ \\
    4.&OD-CLF-CBF-QP & $ \qty{2.09}{sec}$ & $0.0430$ \\
    5.&Optimal Control (OCP) & $ \qty{17.2}{sec}$ & $0.0130$ \\
\bottomrule
\end{tabular}
\end{table}

\cref{fig:Sim} shows that all controllers successfully perform rest-to-rest maneuvers despite large initial angles. The RES-CLF-QP experiences input chattering, while the OD-CLF-QP and OD-CLF-CBF-QP maintain smooth performance due to the varying decay weight $\rho$ (\cref{fig:rho}) described in \cref{sec:chatter}. Chatter can also be eliminated by increasing the sampling rate, but the required increase is a factor of 100, which is an untenable computational burden.

The Saturated PD, RES-CLF-QP, and OD-CLF-QP, which lack the ability to enforce state constraints, fail to satisfy the RW angular momentum limits (see $h_{w3}$ plot in \cref{fig:Sim}).

As shown in \cref{table:Performance comparison}, the proposed OD-CLF-CBF-QP achieves a cost comparable to the OCP while offering significant computational time savings. Controllers \ref{c2}--\ref{c4} each solve the same number of QPs, but OD-CLF-CBF-QP takes slightly longer since its associated QP has an additional CBF constraint and decision variable $\rho$.

\subsection{Monte Carlo Simulations}

We ran Monte Carlo simulations with 20 random initial orientations (uniformly distributed on a sphere) to demonstrate the effectiveness of the proposed OD-CLF-CBF-QP controller \eqref{eq:Optimal Decay CLF-QP3}. We used the same control parameters as in \cref{sec:comparative simulations}, except we set $\alpha = 1$ to allow the trajectories to approach the state constraints more closely.

\begin{figure}[htp]
	\centering
	\includegraphics{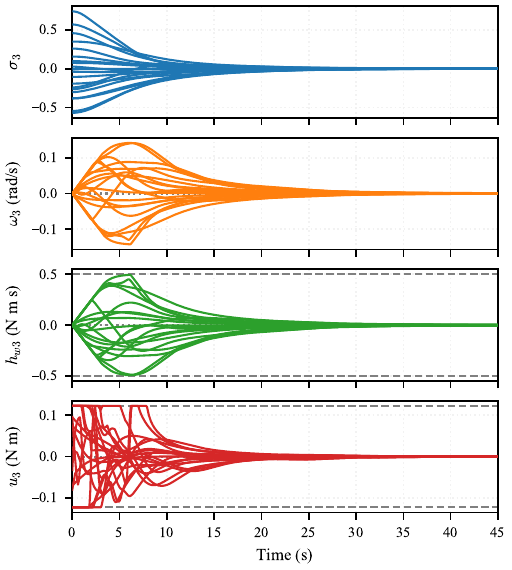}\vspace{-1mm}
	\caption{Simulated representative state and input responses for the OD-CLF-CBF-QP controller for 20 randomized initial orientations.
 Only the 3\textsuperscript{rd} components of $\sigma$, $\omega$, $h_w$, $u$ are shown.}\label{fig:MonteCarlo}
\end{figure}

The simulation results, shown in \cref{fig:MonteCarlo}, verify that the proposed OD-CLF-CBF-QP effectively performs safe rest-to-rest maneuvers with random large initial states while avoiding input chatter.
Code that reproduces all the simulations and figures in this paper is available at \\\url{https://github.com/QCGroup/odclfcbf}.

\section{Conclusion}\label{sec:Conclusion and Future Work}

We proposed a control strategy for agile spacecraft attitude control using control barrier functions (CBF) and quadratic programming (QP) to address reaction wheel torque and angular momentum constraints. The approach introduces a dynamic optimal-decay weight for the control Lyapunov function (CLF) to reduce chattering at low sampling frequencies while balancing control effort and settling time. %Comparative simulations with optimal control solutions validate its efficiency and effectiveness.

\bibliographystyle{IEEEtran}
\bibliography{bibabbrv}

\end{document}